\documentclass[twocolumn]{aastex62}

\usepackage{hyperref}

\def \src {\mbox{A0538$-$66}}
\def \xmm {\emph{XMM-Newton}}

\defcitealias{Davies81}{DP81} 

\accepted{for publication in ApJ Letters}

\shorttitle{Awakening of the fast-spinning  accreting Be/X-ray pulsar A0538-66}
\shortauthors{L. Ducci, S. Mereghetti, and A. Santangelo}

\begin{document}

\title{Awakening of the fast-spinning  accreting Be/X-ray pulsar A0538-66\footnote{Based on observations  obtained with \xmm, an ESA science  mission  with  instruments  and  contributions  directly  funded  by ESA Member States and NASA.}}

\correspondingauthor{Lorenzo Ducci}
\email{ducci@astro.uni-tuebingen.de}

\author[0000-0002-9989-538X]{Lorenzo Ducci}
\affil{Institut f\"ur Astronomie und Astrophysik, Kepler Center for Astro and Particle Physics, Eberhard Karls Universit\"at, Sand 1, 72076 T\"ubingen, Germany}

\author{Sandro Mereghetti}
\affiliation{INAF -- Istituto di Astrofisica Spaziale e Fisica Cosmica, Via A. Corti 12, 20133 Milano, Italy}

\author{Andrea Santangelo}
\affil{Institut f\"ur Astronomie und Astrophysik, Kepler Center for Astro and Particle Physics, Eberhard Karls Universit\"at, Sand 1, 72076 T\"ubingen, Germany}



\begin{abstract}
\src\ is a Be/X-ray binary (Be/XRB) hosting a 69\,ms pulsar.
It emitted bright X-ray outbursts with
peak luminosity up to  $\sim 10^{39}$\,erg\,s$^{-1}$ during the first years after its discovery in 1977.
Since then, it was always seen in quiescence 
or during outbursts with $L_{\rm x}  \lesssim 4 \times 10^{37}$\,erg\,s$^{-1}$.
In 2018 we carried out   \xmm\ observations of \src\
during three consecutive orbits when the pulsar was close to periastron. 
In the  first two observations we discovered a remarkable variability, with  flares of typical durations between $\sim$2$-$50\,s and peak luminosities up to 
$\sim 4\times 10^{38}$\,erg\,s$^{-1}$ (0.2$-$10\,keV). Between the flares the  luminosity was
$\sim 2\times 10^{35}$\,erg\,s$^{-1}$. The flares were absent in the third observation, during which \src\  had a steady  luminosity of
$2\times 10^{34}$\,erg\,s$^{-1}$.
In all observations, the X-ray spectra consist of a softer component, well described by an absorbed power law with photon index $\Gamma_1\approx 2-4$ and  $N_{\rm H}\approx 10^{21} $\,cm$^{-2}$, plus a  harder power-law component ($\Gamma_2\approx 0-0.5$) dominating above $\sim$2 keV.
The softer component shows larger flux variations than the harder one, and a moderate   hardening correlated with the luminosity. 
The  fast flaring activity seen in these observations was never observed before in \src, nor, to our best knowledge, in other Be/XRBs. 
We explore the possibility that during our observations the source was accreting 
in a regime of nearly spherically symmetric inflow.
In this case, an atmosphere can form around the neutron star magnetosphere and the  observed variability can be explained by transitions between the accretion and supersonic propeller regimes.  
\end{abstract}



\section{Introduction} \label{sec:intro}

Be/X-ray binaries (Be/XRBs) consist of a Be star
and, usually, a neutron star (NS). Most of them show
a weak persistent X-ray emission ($L_{\rm X}\lesssim 10^{34}$\,erg\,s$^{-1}$),
interrupted by outbursts ($L_{\rm X}\lesssim 10^{38}$\,erg\,s$^{-1}$) which
last several weeks. The outbursts are caused by accretion onto the NS of the plasma captured from the circumstellar disks that characterize Be stars
(for a review see, e.g., \citealt{Reig11}).

\src\ is a Be/XRB located in the Large Magellanic Cloud (LMC). It hosts one of the fastest
spinning pulsars (period $P$= 69\,ms)  and has one of the shortest  orbital periods  
($P_{\rm orb}=16.6409\pm 0.0003$\,d) and highest eccentricities ($e=0.72$)  among Be/XRBs \citep{Rajo17,White78}.
These characteristics might be at the basis of the peculiar properties observed in this system, both in X-rays and in the optical band\footnote{For the peculiar optical properties shown by \src, see \citet{Ducci19,Ducci16} and references therein.}.
The outbursts observed in the first years  after its discovery exceeded the   isotropic Eddington limit, reaching  peak X-ray luminosities of $L_{\rm x}\gtrsim 8\times 10^{38}$\,erg s$^{-1}$
\citep{White78,Johnston79,Johnston80,Skinner80,Ponman84,Skinner82},
while all the subsequent observations caught \src\ at lower X-ray luminosities, in the range
$L_{\rm x}\approx 5\times 10^{33}-4\times10^{37}$\,erg\,s$^{-1}$ 
\citep{Mavromatakis93,Campana97,Campana02,Corbet97,Kretschmar04}.

Remarkably, the pulsations at 69 ms were detected only once,  during a bright outburst
($L_{\rm x} \approx 8\times 10^{38}$\,erg\,s$^{-1}$, \citealt{Skinner82}) observed by the \emph{Einstein} satellite in 1980. They were never detected in all the subsequent observations, either in quiescence ($L_{\rm x} \lesssim 10^{34}$\,erg\,s$^{-1}$) or in outbursts that reached lower luminosities  ($L_{\rm x} \lesssim 10^{38}$\,erg\,s$^{-1}$).
This led to the suggestion that the accreting plasma could overcome the centrifugal magnetospheric barrier and  reach the NS surface, thus producing X-ray pulsations,  only during episodes of very high accretion rate
\citep{Campana95,Corbet97}.

In fact, if the rate of mass gravitationally captured by a NS 
is below a minimum value, 
that depends on the magnetic field strength and the spin period of the pulsar, 
the NS magnetosphere 
is larger than the corotation radius r$_{\rm co} = [GM_{\rm ns} P^2/(4\pi^2)]^{1/3}$ (the distance  at which 
a test particle in a Keplerian orbit corotates with a NS of mass $M_{\rm ns}$ and spin period $P$).
When this occurs, the matter flow is halted at the magnetospheric radius r$_{\rm m}$ and, assuming that all the potential energy of the mass inflow is converted to radiation, 
the X-ray luminosity is reduced by a factor r$_{\rm m}/$R$_{\rm ns}$,   where  R$_{\rm ns}$ is the NS radius. 
Based on these considerations, \citet{Skinner82} and
\citet{Campana95} estimated for \src\ 
an upper limit for the
magnetic dipole moment of $\mu \lesssim 10^{29}$\,G\,cm$^3$. 

In this Letter we report the results of  new  \xmm\ observations
showing a remarkable variability on short timescales, never observed before in   
\src\ and in other Be/XRBs. 
Such a renewed X-ray activity from \src\ possibly preludes to a reactivation of the super-Eddington
regime that characterized this source during the first years after its discovery.

\begin{table}
\caption{Summary of the \xmm\ observations. \label{tab:table log}}
\resizebox{\columnwidth}{!}{
\begin{tabular}{lcccc}
\hline
\hline
   Name   &   Start time   &   Net exposure  &  $\phi_{\rm start}^a$  &  $\phi_{\rm stop}^a$  \\
          &     (UTC)      & time  (ks)      &                     &                     \\
\hline
obs. A & 2018-05-15 06:04:50 & 9.9 & $-$0.0039 & 0.0091 \\ 
obs. B & 2018-05-31 22:04:38 & 12.0 & $-$0.0026 & 0.0077 \\
obs. C & 2018-06-17 12:34:10 & 12.5 & $-$0.0047 & 0.0053 \\
\hline
\end{tabular}
}
  \begin{list}{}{} 
  \item[\scriptsize $^{\mathrm{a}}$] {\scriptsize  Orbital phase based on the ephemeris of
    \citet{Rajo17}. The phase zero of these ephemeris 
    precedes the periastron by $\Delta \phi = 0.038$.}
  \end{list} 
\end{table}

\section{Observations and data analysis} \label{sec:data analysis}

We observed   \src\  with \xmm\   during  three consecutive orbits in 2018.
The observations were done at   orbital phases close to periastron (see Table \ref{tab:table log}).
Data collected by the European Photon Imaging Camera (EPIC) 
were analysed with the standard Science Analysis System (SAS), version 17.0.0.
Observation data files (ODFs) were processed to produce calibrated
event lists for pn, MOS1, and MOS2, using the {\tt epproc} and
{\tt emproc} tasks. For the pn, single- and double-pixel events
(PATTERN$\leq$4) were used, while for the MOS data, single- to
quadruple-pixel events (PATTERN$\leq$12) were used.
Time intervals affected by high background were identified
and excluded\footnote{See the \xmm\ thread: 
\url{https://www.cosmos.esa.int/web/xmm-newton/sas-thread-epic-filterbackground}}, resulting in  the net exposure times indicated in Table 1.
Source events were extracted from a circular  region centered
at the J2000 coordinates R.A.=\,05:35:41.3, Dec.=\,$-$66:51:51, 
with an ``optimal'' extraction radius of 27\,arcsec for obs.\,A
and 29\,arcsec for obs.\,B. These radii were calculated with the SAS task {\tt eregionanalyse} to have the maximum signal to noise ratio.
During obs.\,C, \src\ 
had a much smaller flux than in obs.\,A and B,  but it was still detected
with  high significance (detection likelihood $L=47.69$, corresponding to
spurious probability  $p\approx 2\times 10^{-21}$;
see \citealt{Ducci13} for the source detection procedure adopted here). For this observation, we used a source extraction radius of 20\,arcsec.
The background was extracted from  source-free circular regions.
The times of the events were corrected to the solar system
barycenter with the {\tt barycen} task.

For each observation, we extracted pn lightcurves with binsize of 1\,s,
background subtracted, and corrected for vignetting, bad pixels,
PSF variations, and quantum efficiency, using the SAS task {\tt epiclccorr}.
\src\ showed a strong flux variability (see Sect. \ref{sec:results})
and it was affected by pile-up during the high luminosity states.
For the pn, we generated a response file that includes pile-up corrections\footnote{We 
followed the procedure described in the SAS thread: 
\url{https://www.cosmos.esa.int/web/xmm-newton/sas-thread-epatplot}}.
We verified the goodness of the resulting  spectrum by comparing it with that obtained using the standard response file
and excising the core of the PSF.
Since a response file   including pile-up corrections
cannot be produced for the MOS,   pile-up effects from these data can be removed
only by excising the core of the PSF, which leads to a substantial 
reduction of the statistics. Therefore, in the following analysis we considered only the
pn data for the high and intermediate luminosity levels, while we merged pn and MOS data for the
low luminosity level (see Sect. \ref{sec:results} for the definition of the luminosity levels).

Timing and spectral analyses were performed using 
the standard tools available within HEASOFT v.\,6.24
including {\tt xspec} (v. 12.10.0c; \citealt{Arnaud96}).
For the interstellar absorption, we used the  {\tt tbvarabs} model
with the \citet{Wilms00} abundances  and the photoionization cross-sections of \citet{Verner96}.
\src\ is located in the LMC, an environment with a very different metallicity compared
to the Interstellar Medium (ISM) of the Galaxy \citep{Zhukovska13,Russell92}.
Therefore, we set the following abundances (with respect to the ISM): O: 0.33; Ne: 0.41; Na: 0.45; Mg: 0.48; Si: 0.59; S: 0.48; Fe: 0.38 \citep{Hughes98,Andrievsky01}.
For the other elements heavier than oxygen, we assumed relative abundances
of 0.4 and we left the default values for the other parameters.
We noted that also the simplest model {\tt tbfeo} gives acceptable results,
though with $\chi^2$ values slightly worse than those obtained with {\tt tbvarabs}.

In the following we assume for \src\ a distance of  $d=50$\,kpc \citep{Alves04}.

\section{Results} \label{sec:results}

\begin{figure*}[ht!]
\includegraphics[width=6cm]{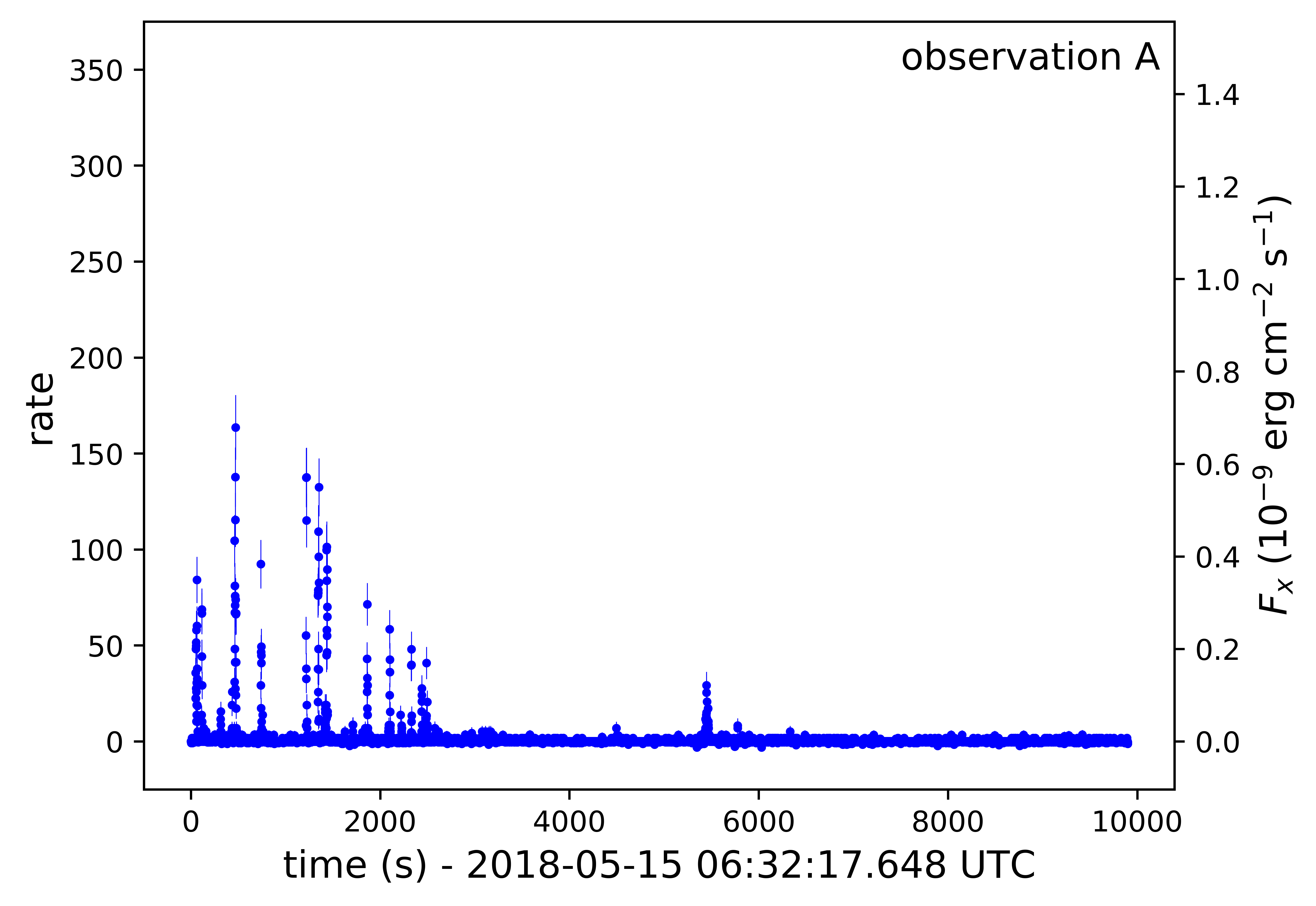}
\includegraphics[width=6cm]{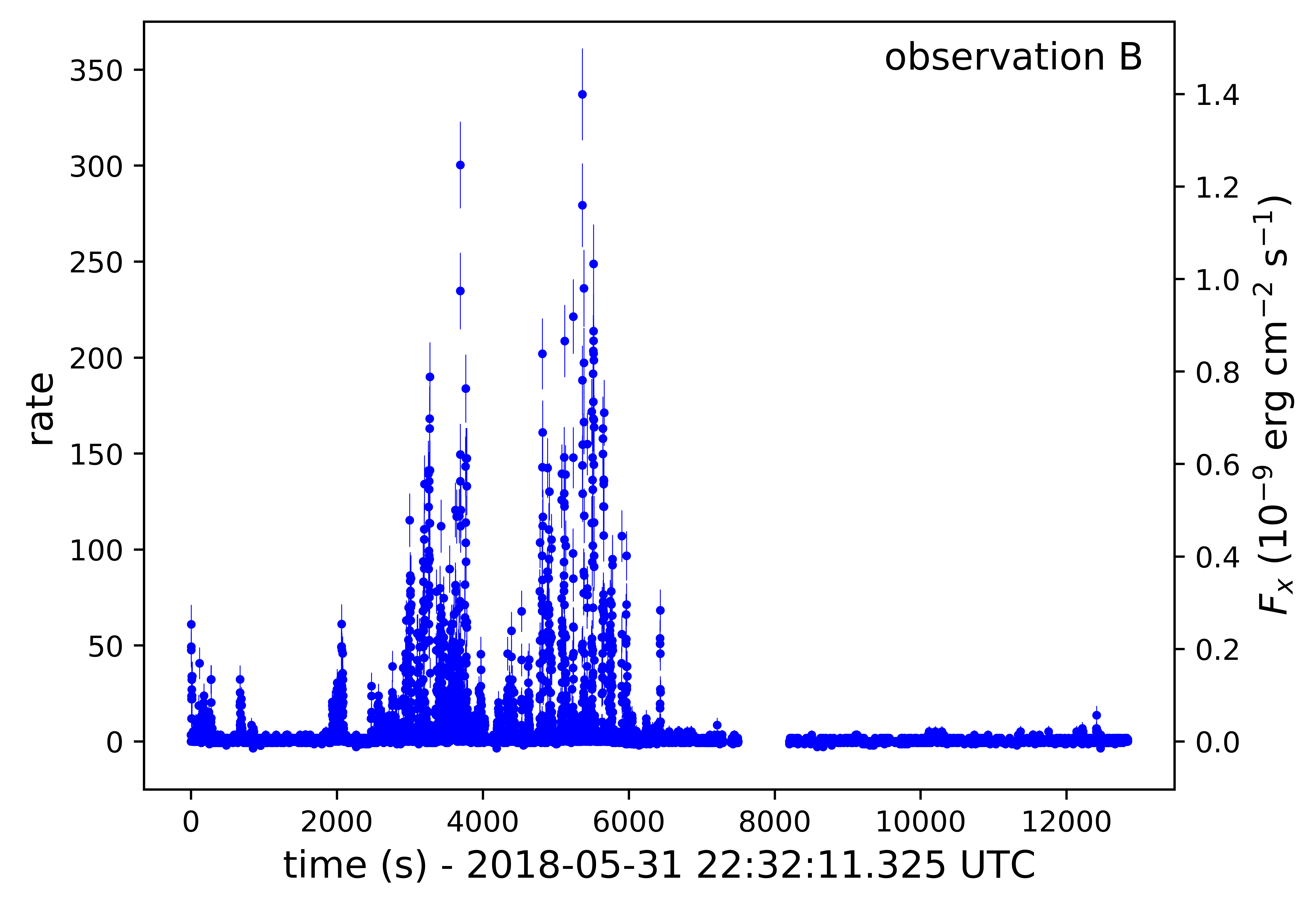}
\includegraphics[width=6cm]{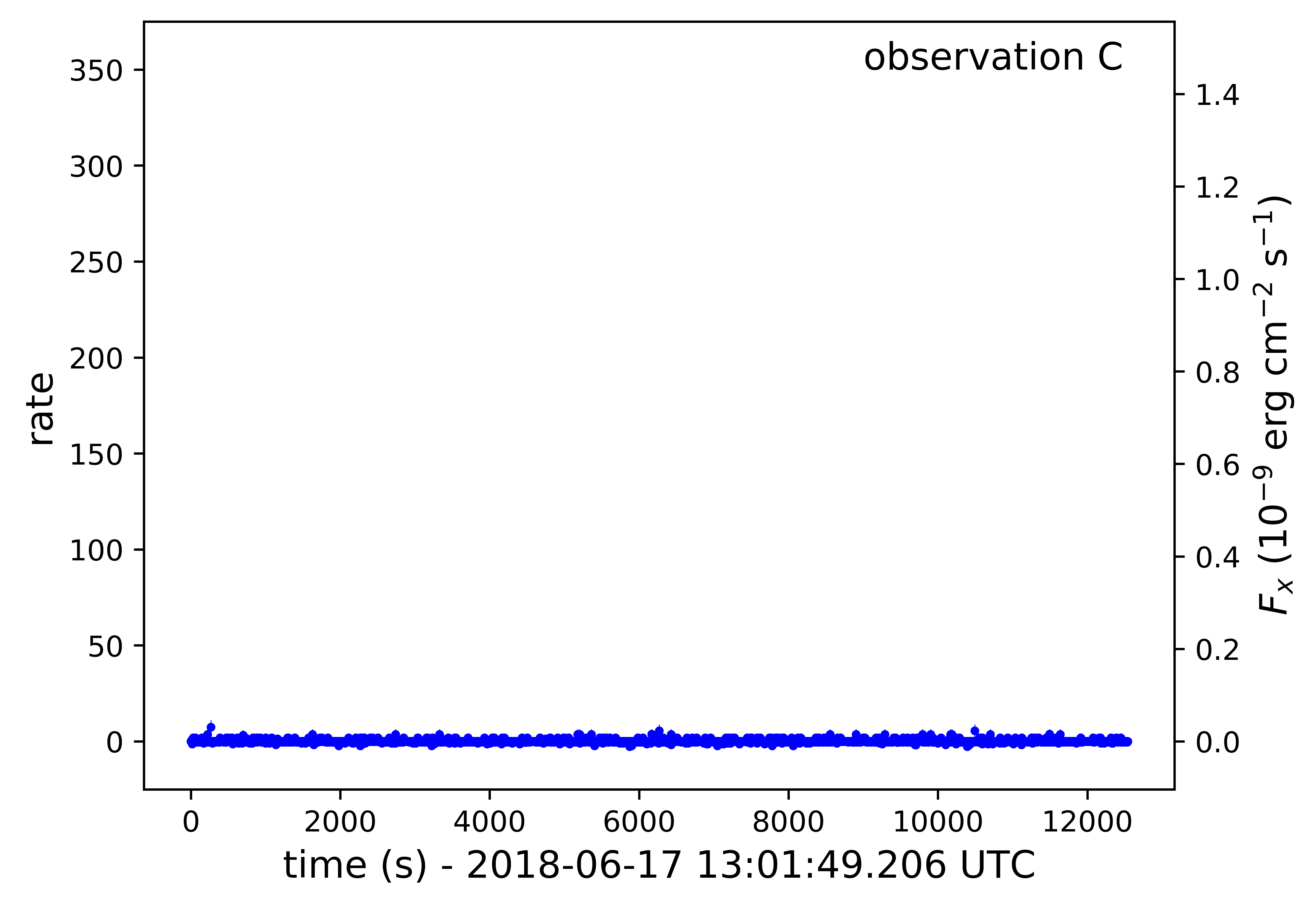}
\includegraphics[height=4.4cm]{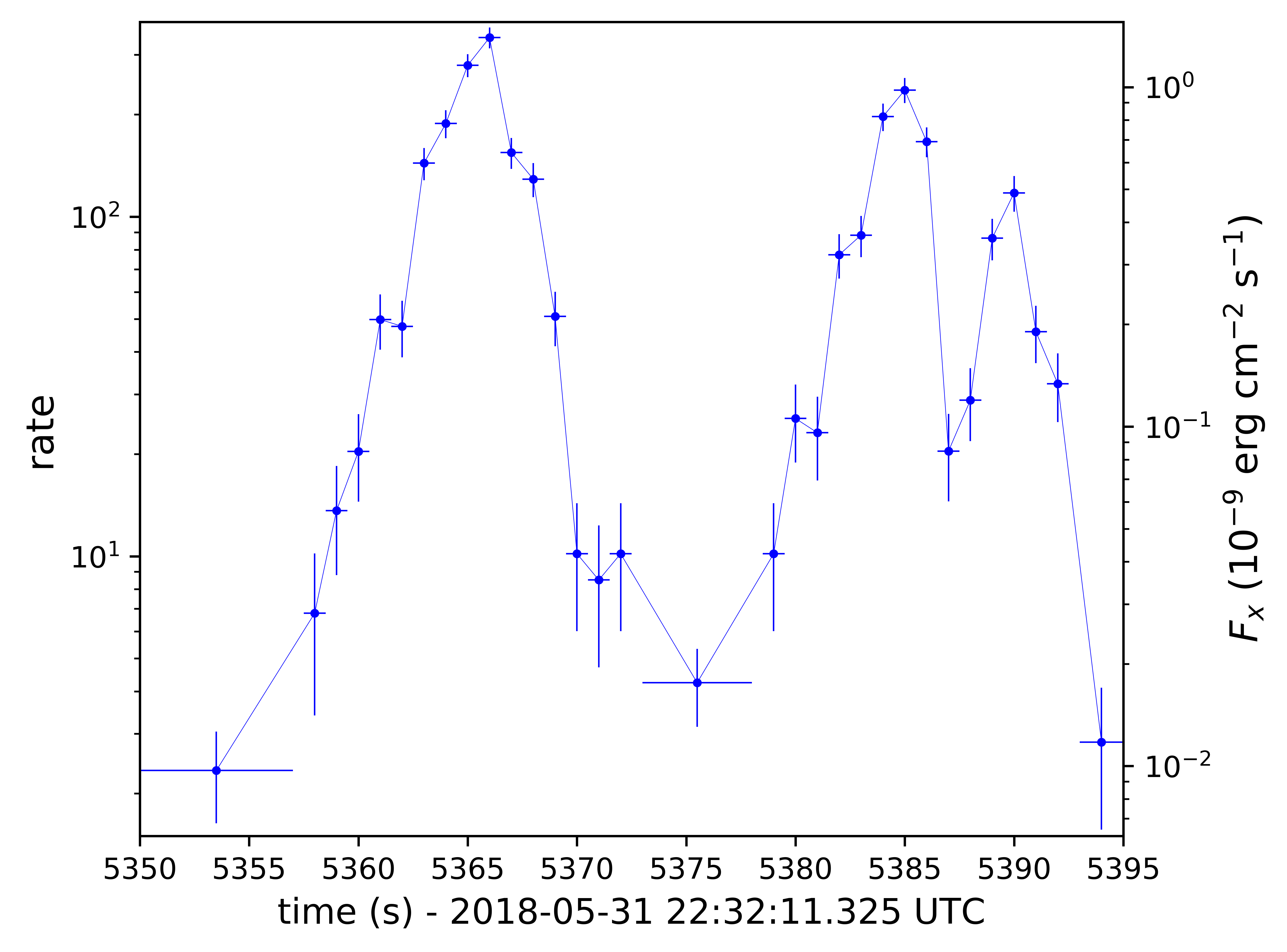}
\includegraphics[height=4.4cm]{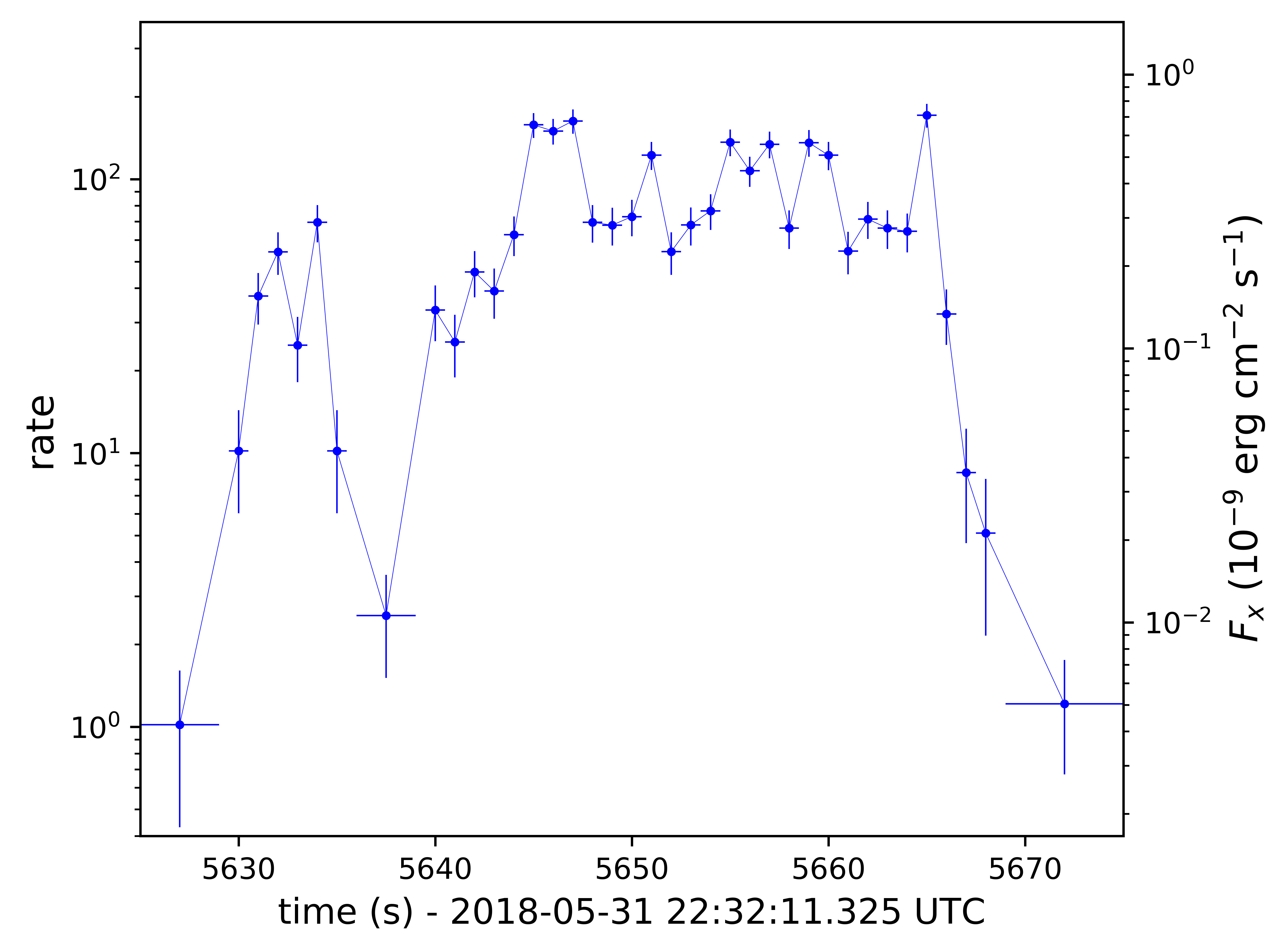}
\includegraphics[height=4.6cm]{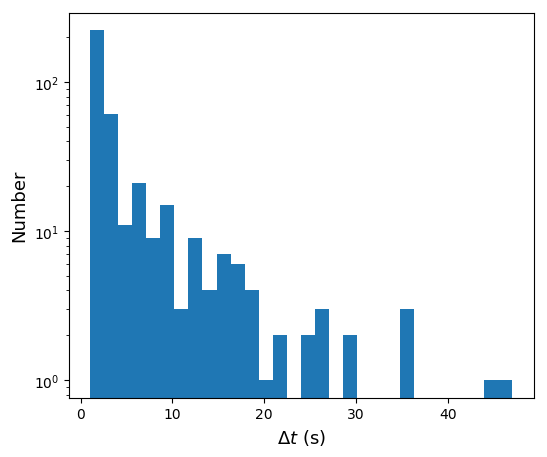}
\caption{Top panel: pn lightcurves (0.2$-$12\,keV, bin time 1\,s) of \src\ during the three   observations.
         Bottom-left panels: two zoomed in sections of the lightcurve of observation B, rebinned at low rates, to better show the structures of the flares. Bottom-right panel: distribution of the durations of the flares ($\Delta t$).
         \label{fig:lcrs}}
\end{figure*}

The X-ray lightcurves (1\,s bin) of \src\ obtained in the three  observations are shown in Fig. \ref{fig:lcrs}.
During the first two observations (A, B) the source was in a very peculiar state of rapid variability,
characterized by very short flares
spanning more than three orders of magnitude, from 
$F_{\rm min\,A,\,B}\approx 5.7\times 10^{-13}$\,erg\,cm$^{-2}$\,s$^{-1}$ 
to $F_{\rm max\,A,\,B}\approx 1.4\times 10^{-9}$\,erg\,cm$^{-2}$\,s$^{-1}$ (0.2$-$12\,keV).
These fluxes correspond to luminosities of
$L_{\rm min\,A,\,B}\approx 1.7\times 10^{35}$\,erg\,s$^{-1}$ and
$L_{\rm max\,A,\,B}\approx 4.2\times 10^{38}$\,erg\,s$^{-1}$.
The distribution of flare durations shows the presence of a large number of  flares shorter than a few seconds (see Fig. \ref{fig:lcrs}).
During  observation C, the source flux was stable and much lower than in the previous two observations: 
$F_{\rm C}\approx 7\times 10^{-14}$\,erg\,cm$^{-2}$\,s$^{-1}$  (0.2$-$12\,keV),
that corresponds to $L_{\rm C}\approx 2.1 \times 10^{34}$\,erg\,s$^{-1}$.
Note that the average luminosity during the ``non-flaring'' time intervals of observations A and B was about eight times higher than $L_{\rm C}$.

We searched for periodic modulations in  the 0.2$-$12\,keV pn events using a Rayleigh test $Z^2$ (e.g. \citealt{Buccheri83}).
The search was limited to periods longer than 12 ms by the
time resolution of the pn camera in small-window mode. 
No statistically significant pulsations were detected.
We calculated the 3$\sigma$ upper-limit on the pulsed fraction $p_{\rm f}$
(defined as the ratio between the difference and sum of the maximum and minimum
count rates of the pulse profile) using the method described in \citet{Brazier94},
in the period range $\sim 50-100$\,ms (including the value of $\sim 69$\,ms
discovered by \citealt{Skinner82}). We found: obs.\,A: $p_{\rm f} \leq 15$\%;
obs.\,B: $p_{\rm f} \leq 9$\%; obs.\,C: $p_{\rm f} \leq 76$\%.
The pulsed fraction of \src\ measured by \citet{Skinner82} for the unique detection
of pulsation from this source was $\sim 26$\%.

To search for possible spectral variability as a function of the X-ray luminosity
we divided the data in three subsets based on the values of the pn count rate:   low (rate $< 5$\,c\,s$^{-1}$)
intermediate ($5\leq$rate$\leq 80$\,c\,s$^{-1}$), and high
(rate$> 80$\,c\,s$^{-1}$).
The boundary   between the intermediate and  high level was chosen to have approximately  the same
statistics in both data sets.
After checking that the pn and MOS spectra for the low state gave consistent results, 
we combined them using the SAS task {\tt epicspeccombine}.
We used a similar procedure to combine the pn spectra of  observations A and B for the intermediate and high levels.

We fitted these spectra in the 0.2$-$12 keV energy range. Using simple single-component models we could not obtain good fits, because the spectra clearly show two distinct components in the soft ($\lesssim 2$\,keV) and hard energy range.
In the following, we concentrate on the simplest phenomenological model that gave a reasonably good fit, i.e. the sum of two absorbed power laws (with the addition of a broad line at $\sim$6.4 keV in the   high and intermediate level spectra).

\begin{figure}[ht!]
  \includegraphics[width=\columnwidth]{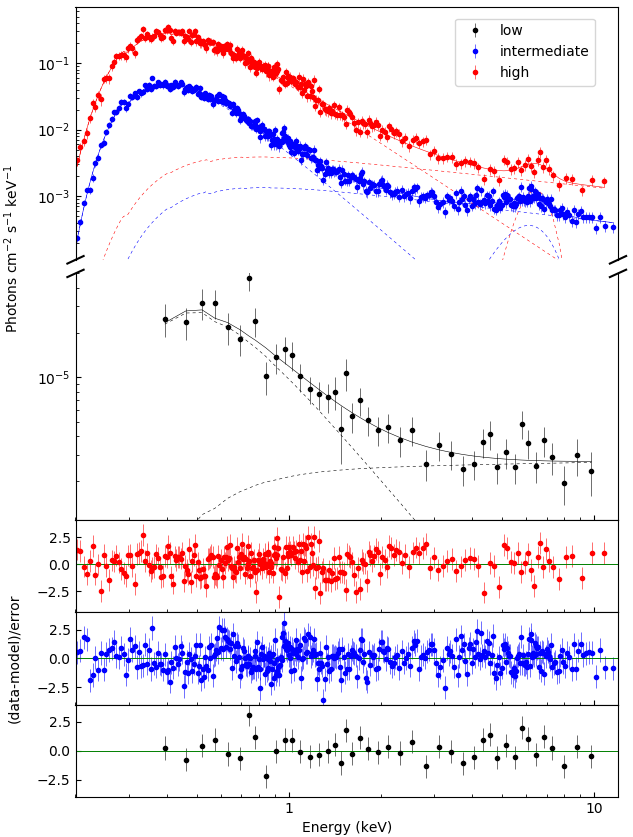}
\caption{\xmm\ spectra of \src\ during the three luminosity levels, fitted with two  absorbed  
power laws (plus a Gaussian line for the intermediate and high  luminosity levels). The lower panels show the residuals of the fits.\label{fig:spectra}}
\end{figure}

The best fit parameters are reported in Table \ref{tab:spectra1} and the corresponding spectra and residuals 
are shown in Fig. \ref{fig:spectra}.
Since the column density is similar in  the three spectra, we also tried to fit them fixing $N_{\rm H}$ to a common value. This led to similar best fit parameters for the power laws, but with worse of chi-squared values.

The comparison of the best fit parameters for the three states  indicates a  moderate spectral variability as a function of luminosity.
In particular, between the intermediate and high level, 
the flux of the softer component increases by a larger factor ($\sim$6) than that of the  harder one ($\sim$3). At the same time,  the low-energy power law becomes harder.

The   intermediate and high level spectra show a broad emission feature 
with energy consistent  
with the K$\alpha$ emission at $6.4$\,keV from \ion{Fe}{23}.
We tried to fit this feature with reflection disk models like 
{\tt diskline}, but this resulted in   worse fits than 
those obtained with a  Gaussian profile.

\begin{deluxetable}{lccc}
\tablecaption{Best fit spectral parameters of the absorbed two-component power law
              plus a gaussian model to describe the three luminosity states of \src\ (errors at 1$\sigma$ confidence level). \label{tab:spectra1}}
\tablehead{
\colhead{Parameters\tablenotemark{a}} & \colhead{low} & \colhead{intermediate} & \colhead{high}   \\
}
\startdata
$N_{\rm H}$ ($10^{22}$\,cm$^{-2}$) &       $0.13{+0.09 \atop -0.07}$    & 
                                        $0.119 \pm 0.006$             & 
                                        $0.098{+0.007 \atop -0.007}$  \\ 
$\Gamma_1$                       &    $2.4{+0.7 \atop -0.6}$ &
                                     $4.0 \pm 0.1$           & 
                                     $3.04{+0.10 \atop -0.09}$ \\
Flux$_1$                        & $0.054{+0.016 \atop -0.009}$ &
                                 $43.3{+2.3 \atop -2.0}$ & 
                                 $272.8{+7.4 \atop -6.9}$ \\
$\Gamma_2$                      & $-0.04{+0.26 \atop -0.35}$ &
                                  $0.54 \pm 0.05$ & 
                                 $0.49 \pm 0.16$  \\
Flux$_2$                        & $0.21{+0.02\atop-0.03}$ & 
                                 $46.8{+1.8\atop -1.9}$ & 
                                 $143.6{+10.2 \atop -10.6}$ \\ 
$E_{\rm line}$   (keV)             &              $-$        &
                                   $6.12 \pm 0.11$ & 
                                   $6.45{+0.16 \atop -0.15}$ \\
$\sigma$ (keV)                  &        $-$       &
                                  $1.02{+0.19 \atop -0.17}$ & 
                                  $0.65 \pm 0.13$ \\
norm line                       &           $-$         &
                                  $9.3{+1.8 \atop -1.5} \times 10^{-4}$ &  
                                 $2.00{+0.49 \atop -0.46} \times 10^{-3}$ \\
$\chi^2$ (d.o.f.)               & 1.107 (39)   &
                                  1.0944 (376) & 
                                  1.2166 (259) \\
norm$_2$/norm$_1$               &  $3.89{+0.31\atop -0.22}$ &
                                 $1.08{+0.07\atop -0.06}$ & 
                                 $0.52 \pm 0.08$ \\
\enddata
\tablenotetext{a}{model tbvarabs*(pegpwrlw+pegpwrlw+gaus) in XSPEC.}
\tablecomments{Unabsorbed fluxes in units $10^{-12}$\,erg\,cm$^{-2}$\,s$^{-1}$ (0.3$-$10\,keV).}
\end{deluxetable}

\section{Discussion} \label{sect. discussion}

The flaring variability  detected in observations A and B, characterized by  flux changes as large as  three orders of magnitude on  timescales 
of a few seconds was never observed before in \src, nor in other Be/XRBs.
Flaring activity has been observed   in a few other high-mass X-ray binaries (HMXBs), but with less extreme properties. For example, the  Be/XRB A0535+26 
showed X-ray flares preceeding an outburst in September 2005 \citep{Caballero08}, but they were much longer ($\Delta t \approx 10^4$\,s), fainter (peak X-ray luminosity of $5\times 10^{36}$\,erg\,s$^{-1}$), and with a smaller dynamic range ($\Delta L\lesssim 10$).
\citet{Postnov08}  explained them as the result of
an interchange instability that develops in the boundary layer between the accretion disk and the NS magnetosphere during the transition from the propeller to the accretion state. 
Similar flares were also observed in another Be/XRBs, EXO\,2030+375, and  explained with an accretion disk-magnetospheric instability, leading to a cyclic increase of the mass accretion rate on the viscous time scale at the magnetosphere \citep{Spruit93,Klochkov11}.

Strong and rapid  variability is also present in the supergiant fast X-ray transients (SFXTs), a subclass of HMXBs with OB supergiant mass donors (see, e.g., \citealt{sidoli2013,Romano15}). Their flares have typical peak luminosity of $10^{36}-10^{37}$\,erg\,s$^{-1}$ (thus 10$-$100  times fainter than those of \src) and    durations of $\sim 10^2-10^3$\,s. 
The mechanism responsible for the flares in SFXTs is not yet clear, although many models involving wind variability, gating mechanisms and settling accretion regimes have been proposed (e.g. \citealt{intZand05,Grebenev07,Bozzo08,Ducci09,Ducci10,Shakura14}).

The flares we observed in \src\  are more reminiscent of those seen
in some accreting millisecond X-ray pulsars (AMXPs, \citealt{Patruno09,Patruno13,Ferrigno14}).
As in some of the models quoted above for other sources, also the AMXPs flares  were explained in terms of magnetic gating mechanisms that can occur in disk-accreting sources   
when $r_{\rm m} \approx r_{\rm co}$ (e.g. \citealt{Spruit93, DAngelo10}).
Notably, also  the AMXPs flares have lower peak luminosities ($\lesssim 10^{36}$\,erg\,s$^{-1}$) and a smaller dynamical range ($\Delta L \approx 10-50$) than those observed in \src.
Another X-ray binary showing similar flares is GRO\,J1744$-$28, also known as the ``Bursting Pulsar''. It consists of a neutron star with spin period of $\sim0.467$\,s accreting from a low mass companion star. It emits type II  bursts, likely caused by viscous instabilities in the accretion disk (see, e.g., \citealt{Bagnoli15} and references therein). 
These bursts have duration of the order of a few seconds and can reach   peak luminosities of $\approx 10^{40}$\,erg\,s$^{-1}$, 
but the amplitude of variability with respect to the non bursting luminosity is of $\Delta L_{\rm x} \approx 6-40$
\citep{Giles96,Sazonov97,Court18}.

\begin{figure}
  \includegraphics[width=\columnwidth]{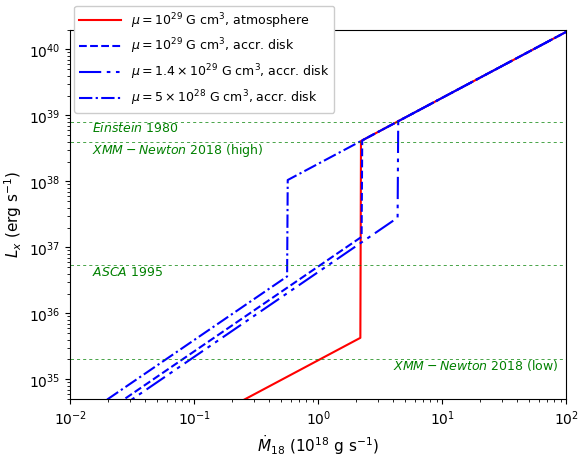}
\caption{Expected luminosity of \src\ as a function of the mass captured rate.
          Blue dashed, dot-dashed, dot-dot-dashed lines show the case of centrifugal inhibition of accretion with an accretion disk, as proposed to explain the previous outbursts of \src, for different values of $\mu$. 
         The  red solid line shows the luminosity regimes for the spherically symmetric accretion scenario of \citetalias{Davies81}.
          Horizontal green dotted lines show the average X-ray luminosities of the 
          most relevant outbursts displayed by \src\ (\emph{Einstein} 1980: \citealt{Skinner82}; 
          \emph{ASCA} 1995: \citealt{Corbet97}).
          \label{fig:lx_vs_mdot}}
\end{figure}

As mentioned above,  
\citet{Campana95} noticed that the presence of pulsations during 
the 1980 super-Eddington flare of \src\  implies an upper limit on its magnetic dipole  $\mu \lesssim 10^{29}$\,G\,cm$^3$.
They  also pointed out that the fainter 
outbursts ($L_{\rm x}\approx 5\times 10^{36}-4\times 10^{37}$\,erg\,s$^{-1}$) seen with \emph{ROSAT} and \emph{ASCA} 
could be explained  with accretion onto the magnetosphere
and that the soft \emph{ROSAT} spectra of the low luminosity states are in agreement with the expected temperature calculated by \citet{Stella94} 
for a standard accretion disk truncated at r$_{\rm m}$.
In this case, assuming that all the potential energy of the accretion flow is released at the magnetosphere  and converted to X-ray radiation,  a luminosity of
$L_{\rm m} \approx G M_{\rm ns}\dot{M}_{\rm c}/r_{\rm m}$ is  produced (see also \citealt{Stella94,King94}).
Given the short spin period of \src, 
a luminosity jump of a factor $\sim$30  (independent on the value of $\mu$) is expected when $r_{\rm m}$ overcomes $r_{\rm co}$ as a result of a decrease of the inflowing mass rate 
\citep{Corbet97}.
This is illustrated in Fig. \ref{fig:lx_vs_mdot}, where the transitions between
the two accretion regimes for different values of $\mu$ are     compared
to the X-ray luminosities of the most relevant X-ray observations of \src.
Clearly, the  luminosity variations seen in the \xmm\ observations reported here are too large to be explained with this
scenario.

In the following, we explore the  possibility that during our observation \src\ was in a regime of spherical accretion and its variability  caused by rapid changes between the different accretion regimes discussed in \citet{Davies81} (hereafter \citetalias{Davies81}).  

An accretion disk can form only if the specific angular momentum of the gravitationally captured matter is sufficiently large. This can be checked by considering the 
circularization radius (see, e.g., \citealt{FKR}), that in case of wind accretion can be estimated as 
\begin{equation} \label{rcirc}
r_{\rm circ} = \xi G^3 M_{\rm ns}^3 \omega^2 v_{\rm rel}^{-8} \mbox{ ,}
\end{equation} 
where $G$ is the gravitational constant, $\omega$
is the orbital angular velocity,
$M_{\rm ns}$ is the NS mass\footnote{we take $M_{\rm ns}=1.4\,M_\odot$ in the whole paper.}, and  $v_{\rm rel}$ is the relative velocity between the NS and the wind\footnote{For the calculation of the orbital separation and the relative wind velocity, we followed \citet{Smart65, Waters89}, and \citet{Rajo17} for the parameters of the binary system.}. 
The factor $\xi\sim0.2$ accounts for the reduction in angular momentum due to inhomogeneities in the  wind \citep{Ikhsanov01b}.
Due to the highly eccentric orbit with a large inclination with respect to the equatorial plane of the Be star \citep{Rajo17}, for most of the time the NS is embedded in the fast ($v\gtrsim 500$\,km\,s$^{-1}$) and weak polar wind
of the companion star. Therefore, $r_{\rm circ}\approx 2\times 10^6$\,cm, 
much smaller than the magnetospheric radius ($\gtrsim 10^8$\,cm), and  a disk cannot form. 
A transient accretion disk might form 
when the NS crosses the Be circumstellar disk, where the wind is denser and slower, but also this possibility is uncertain. For a wind velocity law  $v_{\rm w}(r) = v_0 (r/R_{\rm d})^{n-2}$,
with $v_0=5-50$\,km\,s$^{-1}$, $2.5 \leq n \leq 4$ \citep{Waters89}, and $R_{\rm d}= 10$\,$R_\odot$, we estimate 
$v_{\rm rel}$ at periastron in the range $\sim$3.1$-$4.8$\times10^7$\,cm.
By comparing the resulting 
$r_{\rm circ}\approx 0.13-4.1\times 10^8$\,cm 
with the values of $r_{\rm m}\approx 1.1\times 10^8$\,cm discussed below, 
it can be seen that there are regions in the parameter space for which a disk cannot form.
Finally, we   note that the transient nature of an accretion disk or its absence
is also supported by the occasional lack of the \ion{He}{2}\,$\lambda$4686 
emission line at times of outbursts \citep{McGowan03}.
Based on these considerations, we believe that our assumption of (nearly) spherical accretion is not unreasonable and we can apply the framework described by \citetalias{Davies81}.

From the peak luminosity of the flares we  can estimate the rate of
"captured" mass, $\dot{M}_{\rm c}\approx 2\times 10^{18}$\,g\,s$^{-1}$.
If the drops in luminosity between the flares are caused by the sudden activation of the magnetic barrier,  the magnetospheric radius must be close to the  corotation radius 
$r_{\rm co} = 2.8\times 10^7$ cm.
Therefore, using the  canonical definition of r$_{\rm m}$ (see eq. 2.5 in \citetalias{Davies81}),
\begin{equation} \label{r_m_can}
r_{\rm m} \approx  3.6\times 10^7 \dot{M}_{18}^{-2/7} \mu_{29}^{4/7} M_{1.4}^{-1/7} \mbox{ cm ,}
\end{equation}
where 
$\mu_{29} = \mu/(10^{29}$\,G\,cm$^3)$
and $\dot{M}_{18}=\dot{M}_{\rm c}/(10^{18}$\,g\,s$^{-1})$,
setting 
$\dot{M}_{18}= 2$, we find that there is a transition
from accretion to inhibition of accretion when $\mu_{29}\approx 1$\footnote{
We note that for the mass captured rate implied by the X-ray luminosities of the flares,
\src\ could be in the subsonic regime, with the formation of
an adiabatic atmosphere surrounding the magnetosphere when $r_{\rm m}<r_{\rm co}$ (\citetalias{Davies81}).
Although for the value of $\dot{M}_{\rm c}$ mentioned above the adiabatic atmosphere 
would be stable against damping of convective motions caused by bremmstrahlung radiative cooling,
from equation 21 in \citet{Bozzo08}  it can be noted that during the subsonic regime the luminosity produced by the matter entering the magnetosphere through Kelvin Helmholtz instability has the same order of magnitude of the luminosity the pulsar would
have if it accreted on its surface. In this case, the effects of the X-ray radiation coming from the NS on the atmosphere may no longer be negligible and this regime of accretion could therefore be absent.
}.
   
\citetalias{Davies81} showed that, under certain conditions, a quasi-static atmosphere can form around the NS magnetosphere.
The atmosphere is heated by the conversion of rotational energy 
of the  spinning-down NS,  that is transported from the base of the atmosphere outwards, through convective and turbulent motions. 
The atmosphere remains stable if it does not cool down significantly by radiative losses.
When the magnetospheric radius overcomes the corotation radius, the supersonic propeller regime activates.
\citetalias{Davies81} showed that in this case an atmosphere with an effective polytropic index of $n=1/2$
forms around the NS. Its lower boundary  (the magnetospheric radius) moves to:
\begin{equation} \label{r_m_sup}
r_{\rm m,sup} \approx 8\times 10^7 \mu_{29}^{4/9} \dot{M}_{18}^{-2/9} v_8^{-4/9} M_{1.4}^{1/9} \mbox{ cm ,}
\end{equation}
where $v_8 = v_{\rm rel}/(10^8$\,cm\,s$^{-1}) \approx 0.35$ for \src.   
Setting $\dot{M}_{18}=2$ in Eq. \ref{r_m_sup}, we get $r_{\rm m,sup}\approx 1.1\times 10^8$\,cm. 
$r_{\rm m,sup}$ is larger than the magnetospheric radius given by equation \ref{r_m_can}.
\citet{Lipunov87} showed that this   can be qualitatively explained  by the decrease in density and pressure of the atmosphere due to its heating, which causes its expansion.
\citetalias{Davies81}, and later \citet{Ikhsanov02}, showed that the atmosphere in the supersonic propeller regime is stable against bremsstrahlung cooling and does not collapse until the mass captured rate is lower than:
\begin{equation} \label{Mdotlimsup}
\dot{M}_{\rm lim} \approx 3.1 \times 10^{18} M_{1.4} v_8 \mbox{   g\,s}^{-1}  \mbox{ .}
\end{equation}
The exact value of $\dot{M}_{\rm lim}$ is subject to some uncertainties \citep{Bozzo08}.
It is important to note that $\dot{M}_{\rm lim}$ is derived from the mixing length theory of convection, which is a crude simplification of the physical process of convection \citep{Cox68}. 
$\dot{M}_{\rm lim}$ also depends on the detailed derivation presented in different works. If we use 
the treatment of the convective efficiency parameter of \citet{Kippenhahn90} (instead of that of \citealt{Cox68} used by \citealt{Ikhsanov02}), $\dot{M}_{\rm lim}$ would be higher by a factor of two.

The luminosity in the supersonic propeller regime is produced by the conversion of the
rotational energy dissipated at the lower boundary of the atmosphere (\citetalias{Davies81}),
and is given by:
\begin{equation} \label{Lsd}
L_{\rm sd} \approx 8\times 10^{34} \dot{M}_{18} v_8^2 \mbox{ erg\,s}^{-1} \mbox{ .}
\end{equation}
For $\dot{M}_{18}=2$ and $v_8=0.35$, we obtain $L_{\rm sd}\approx 2\times 10^{34}$\,erg\,s$^{-1}$, which is lower than the intra-flare luminosity in the first two \xmm\ observations. 
In addition, we did not observe   strong spectral variations
between the flares and the low-luminosity states, although these could have been  expected in the framework of the scenario of \citetalias{Davies81} (see also \citealt{Ikhsanov01}).
These difficulties can be overcome if we consider the possibility that a fraction
of the material in contact with the magnetosphere leaks towards the NS surface through the magnetospheric barrier via magnetic reconnections.
According to the ``reconnection driven accretion model'' of \citet{Ikhsanov01}
and the work of \citet{Elsner84}, the rate of plasma accreted because of reconnection of the magnetic field lines is:
\begin{equation} \label{Mdotrec}
\dot{M}_{\rm rec} \approx 10^{15} \left [ \frac{\alpha_{\rm R}}{0.1} \right ] \left [ \frac{\lambda_{\rm m}}{0.01 r_{\rm m}} \right ] \dot{M}_{18} \mbox{ g\,s}^{-1} \mbox{ ,}
\end{equation}
where $\alpha_{\rm R}\approx 0.1$ and $\lambda_{\rm m}\approx 0.1-0.01r_{\rm m}$ (\citealt{Ikhsanov01} and references therein).
Using Eq. \ref{Mdotrec}, we find that the luminosity caused by magnetic reconnections in \src\ could be of the order of $L_{\rm x} \approx 10^{35}$\,erg\,s$^{-1}$, in agreement with the observations.
The red solid line of Fig. \ref{fig:lx_vs_mdot} shows the expected X-ray luminosity in this scenario, including  both the contributions of Eqs. \ref{Lsd} and \ref{Mdotrec}.
The instabilities arising around the transition between
accretion and supersonic propeller regime might produce the flares of the \xmm\ observations
presented here.

Finally, we mention a possible qualitative interpretation of the spectral variability observed in our data.
It is based on the possibility that, during the low luminosity levels,
accretion is not completely inhibited by the centrifugal barrier
and   a fraction of the matter can leak from the inner layers of the atmosphere onto the NS surface  (see, e.g., \citealt{Elsner84}).
This is supported by the observation of accretion episodes at luminosities below the
transition limit between the accretion and the centrifugal inhibition regimes observed in other X-ray binaries (e.g. \citealt{Rutledge07,Doroshenko14}).
In the framework of the idea  
proposed by \citet{Zhang98} to explain the hard X-ray spectrum of Aql\,X$-$1,
the soft spectral component of \src\ could be produced by the accretion of matter onto the NS surface.
The hard component is produced by inverse Compton scattering of the photons of the soft component by the electrons in the atmosphere just outside the
magnetosphere during the flares and
the low luminosity states (if magnetic reconnections takes place).
According to the recent findings of \citet{Tsygankov19}, 
  bulk Comptonization of the leaking matter should be negligible
because of the small optical depth expected at the accretion rates occurring
during the low luminosity level of \src.
When the accretion on the surface decreases dramatically, the soft component decreases suddenly.
The hard component also decreases as a result of the decrease of the seed photons. However, according to \citet{Wang85}, the temperature outside the magnetosphere during the supersonic propeller regime can increase and the power law that describes the
hard X-ray emission produced by Comptonization will become harder, similarly to what observed in \src.

\section{Conclusions} \label{sect. conclusions}

Our  new X-ray data  (obtained sixteen years after the last
observation of \src) led to the discovery of a peculiar flaring behavior, never seen before in this source.
Although other explanations for the observed variability cannot be excluded, we  speculate that the strong and rapid flares occur because the source was accreting from a spherically symmetric flow, not mediated by an accretion disk. 
In these conditions an atmosphere can form above the NS magnetosphere and flares might be produced by rapid changes between the accretion and supersonic propeller regime.
On the other hand the  less dramatic variability observed in previous occasions is consistent with  episodes of accretion from a disk.
Both accretion scenarios are possible provided that the  magnetic dipole moment is 
$\mu \approx 10^{29}$\,G\,cm$^3$.
In general, a thorough study of the spectral properties would require a better coverage at higher energies to better constrain the hard component.

\acknowledgments

LD acknowledges the kind hospitality of INAF/IASF-Milano, where part of this work was carried out.
This work is supported by the Bundesministerium
f\"ur  Wirtschaft  und  Technologie  through  the  Deutsches  Zentrum  f\"ur  Luft  und
Raumfahrt  (grant  FKZ  50  OG  1602) and
by the agreement ASI/INAF I/037/12/0.



\bibliographystyle{aasjournal} 
\bibliography{ld}

%



\end{document}